\documentclass[10pt,english]{article}
\usepackage{times}
\usepackage[T1]{fontenc}
\usepackage[latin1]{inputenc}
\usepackage{geometry}
\usepackage{subfigure}
\usepackage{graphicx}
\usepackage{setspace}
\makeatletter

\providecommand{\LyX}{L\kern-.1667em\lower.25em\hbox{Y}\kern-.125emX\@}
\newcommand{\noun}[1]{\textsc{#1}}
 \newcommand{\lyxaddress}[1]{
   \par {\raggedright #1 
   \vspace{1.4em}
   \noindent\par}
 }

\usepackage{babel}
\makeatother
\begin{document}

\title{\noun{Diffraction Anomalous Fine Structure spectroscopy at the
beamline BM2 at the European Synchrotron Radiation Facility}}

\author{H. Renevier$^{1*,2}$, S. Grenier$^{1,3**}$, S. Arnaud$^{1}$, J.F.
Bérar$^{1}$, B. Caillot$^{1}$, \\
J.L. Hodeau$^{1}$, A. Letoublon$^{1*}$, M.G. Proietti$^{4}$, B.
Ravel$^{5}$}

\maketitle

\lyxaddress{$^{1}$Laboratoire de Cristallographie, Centre National de la Recherche
Scientifique, BP 166, F-38042 Grenoble Cedex 09, France.}

\lyxaddress{$^{2}$Université Joseph Fourier, BP 53, F-38041, Grenoble Cedex
09, France. }

\lyxaddress{$^{3}$ Dept. of Physics and Astronomy, Rutgers University, Piscataway,
08854, New Jersey, USA.}

\lyxaddress{$^{4}$Departamento de Física de la Materia Condensada, Instituto
de Ciencia de Materiales de Aragón, CSIC-Universidad de Zaragoza -
c. Pedro Cerbuna 12, 50009 Zaragoza, Spain}

\lyxaddress{$^{5}$ Center for Corrosion Chemistry and Engineering, code 6134,
Naval Research Laboratory, Washington DC 20375, USA.}

\lyxaddress{$^{*}$Present address : Commissariat à l'Energie Atomique, Département
de Recherche Fondamentale sur la Matière Condensée, SP2M/NRS, 17 rue
des Martyrs, 38054 Grenoble Cedex 9, France. Hubert.Renevier@cea.fr}

\lyxaddress{$^{**}$Present address : Dept. of Physics, Brookhaven National Laboratory,
Upton, 11973, USA.}

Diffraction Anomalous Fine Structure (DAFS) spectroscopy uses resonant
elastic x-rays scattering as an atomic, shell and site selective probe
that gives information on the electronic structure and the local atomic
environment as well as on the long range ordered crystallographic
structure. A DAFS experiment consists of measuring the Bragg peak
intensities as a function of the energy of the incoming x-ray beam.
The French CRG (Collaborative Research Group) beamline BM2-D2AM (Diffraction
Diffusion Anomale Multi-longueurs d'onde) at the ESRF (European Synchrotron
Radiation Facility) has developed a state of the art energy scan diffraction
set-up. In this article, we present the requirements for obtaining
reliable DAFS data and report recent technical achievements.

\section{Introduction }

Before entering into the technical details, a discussion of the interest
in DAFS spectroscopy is merited. A DAFS experiment consists of measuring
the elastic scattering intensity as a continuous function of the incoming
x-ray beam energy in regions spanning absorption edges. It provides
information about the chemical state and the local environment of
the resonant atom (also known as the anomalous atom), like x-ray Absorption
Fine Structure (XAFS) spectroscopy. But in contrast to XAFS, it is
a \emph{chemical-selective} and \emph{site-selective} spectroscopy.
Like Multiple-wavelength Anomalous Diffraction (MAD), DAFS provides
with a means to recover the phase of the structure factor, important
for solving the long-range average crystallographic structure.

Observation of x-ray Diffraction Anomalous Fine Structure (DAFS) was
reported for the first time in the mid fifties by Y. Cauchois \cite{Cauchois56a}.
With a dispersive diffraction setup using a wavelength bent single
crystal of mica, Cauchois observed intensity variations of the (002)
reflection close to the Al K-edge and suggested that they could be
anomalous diffraction of the mica crystal. To our knowledge \textbf{}no
\textbf{}further contribution to developing the DAFS technique was
made \textbf{}until \textbf{}the publications of Fukamachi \emph{}et
al\emph{.} in 1977 \cite{Fukamachi77} and Salem \emph{et} al\emph{.}
in 1980 \cite{Salem80}. In 1982 Ardnt et al\emph{.} \cite{Arndt82}
showed the possibility of performing anomalous dispersive diffraction
measurements at synchrotron radiation sources for the phase determination
of the structure factor. In 1987 Arcon et al\emph{.} \cite{Arcon87}
measured with an x-ray tube the Bragg reflectivity extented structure
at the Cu K-edge of a copper sulfate single crystal. The data clearly
exhibited anomalous diffraction oscillations and a method to analyse
them was proposed. Although these were pioneering experiments, attention
was paid neither by the diffraction community nor by the absorption
community to the measurement of diffraction intensity as a function
of the energy near absorption edges. In 1992, Stragier et al\emph{.}
\cite{Stragier92} presented an elegant demonstration of the so-called
DAFS spectroscopy on a copper single crystal. At the same time, I.J.
Pickering et al\emph{.} \cite{Pickering93a,Pickering93b} were measuring
the DAFS spectra of powder Fe$_{3}$O$_{4}$ to extract site selective
spectra of Fe in the octahedral and tetrahedral sites. Later, several
groups applied the method at synchrotron radiation facilities with
both monochromatic and dispersive optics to study thin films, multi-layers,
powders, and single crystals. In this paper, we present the setup
and technical requirements of the DAFS experiment. The reader can
find extensive information about DAFS and anomalous diffraction in
the review articles by Sorensen et al. \cite{Sorensen94} and more
recently by Hodeau et al. \cite{Hodeau2001}. The article by Proietti
\emph{}et al. \cite{Proietti99} develops the analysis of the Extented-DAFS
(EDAFS), i.e. the portion of the DAFS spectrum above the absorption
energy.

\section{Elementary background}

\label{sec:Elementary-background}

\subsection{The complex atomic scattering factor}

\begin{figure}[htbp]
\begin{center}\subfigure[]{\includegraphics[  width=0.45\textwidth,
  height=205pt,
  keepaspectratio]{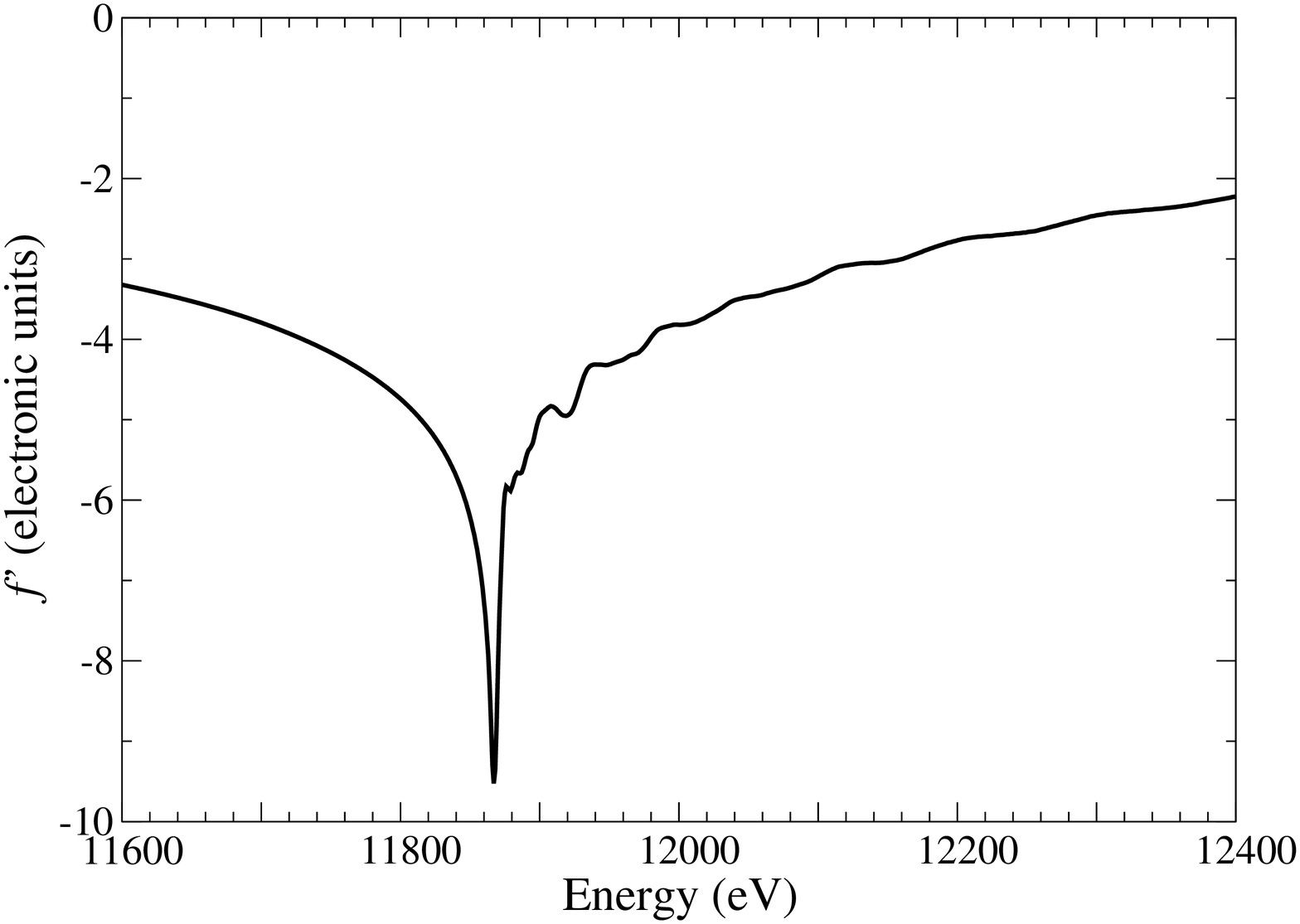}}\subfigure[]{\includegraphics[  width=0.45\textwidth,
  keepaspectratio]{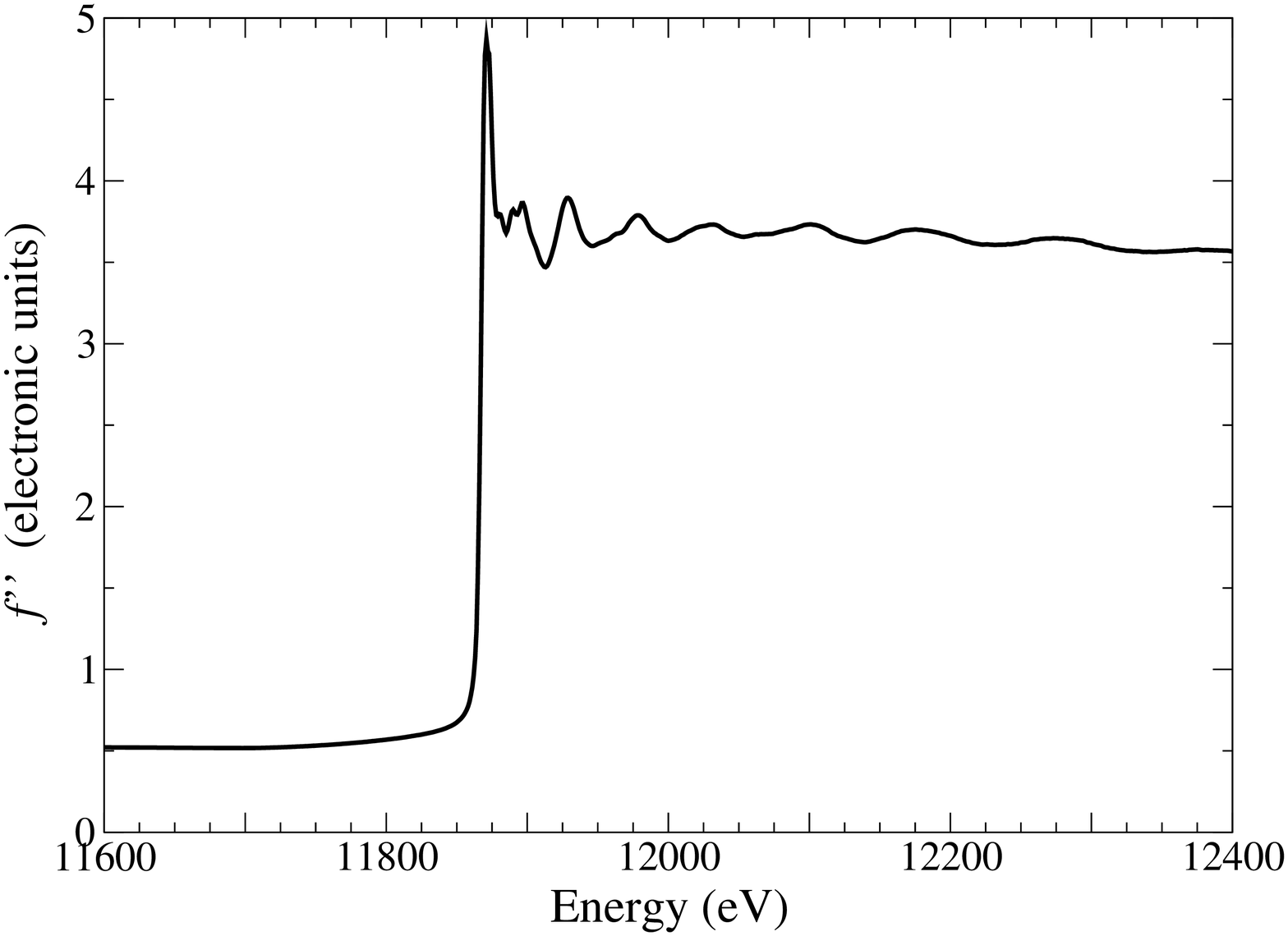}}\end{center}

\caption{\label{fig:fpsfs_InAs} Experimentally determined (a) $f_{As}^{\prime }$
and (b) $f_{As}^{\prime \prime }$, the real and imaginary resonant
scattering of As atoms in InAs bulk, respectively. The $f_{As}^{\prime \prime }$
spectrum was obtained from an absorption measurement of an InAs powder
using Equation \ref{mathed:f''A(sigma)}, $f_{As}^{\prime }$ is the
Kramers-Kronig transform of $f_{As}^{\prime \prime }$ calculated
with the program DIFFKK \cite{Cross98b}. The fine structure oscillations
give information about the local atomic environment of the resonant
atom.}
\end{figure}

Consider the scattering of an atom \emph{A} in a solid sample. Near
an absorption edge the elastic scattering of that atom is expressed
as the sum of the energy independent Thomson scattering $f_{0A}(\overrightarrow{Q})$
and the complex resonant (anomalous) scattering : \begin{equation}
f_{A}(\overrightarrow{Q},E)=\hat{\varepsilon }.\hat{\varepsilon }^{\prime }f_{0A}(\overrightarrow{Q})+f_{A}^{\prime }(E,\vec{k},\vec{k}^{\prime },\hat{\varepsilon },\hat{\varepsilon }^{\prime })+if_{A}^{\prime \prime }(E,\vec{k},\vec{k}^{\prime },\hat{\varepsilon },\hat{\varepsilon }^{\prime })\label{eq:fA}\end{equation}
where $\vec{Q}=\vec{k}^{\prime }-\vec{k}$ is the scattering vector,
$\vec{k}$ and $\vec{k}^{\prime }$ the incident and outgoing wave
vectors, $\hat{\varepsilon }$ and $\hat{\varepsilon }^{\prime }$
the polarization vectors of the incident and outgoing beams (unitary
vectors), respectively \cite{Sakurai67,Blume85,CohenPhotonsAtomes}.
Within the dipole-dipole approximation the resonant scattering depends
only on polarizations and energy. As a general rule, the complex resonant
scattering is not a scalar and can be expressed as $f_{A}^{\prime }(E,\hat{\varepsilon },\hat{\varepsilon }^{\prime })+if_{A}^{\prime \prime }(E,\hat{\varepsilon },\hat{\varepsilon }^{\prime })=\hat{\varepsilon }^{\prime }\left[D\right]\hat{\varepsilon }$,
where $\left[D\right]$ is a tensor of rank 2, whose symmetry is given
by the point group symmetry of the crystallographic site A  \cite{Templeton80,Dmitrienko83}.
In the forward scattering limit, the optical theorem \cite{CohenMecaQII}
shows that the imaginary part of the elastic resonant scattering is
proportionnal to the total cross section : \begin{equation}
f_{A}^{\prime \prime }(\vec{Q}=0,E)=\frac{E}{2hcr_{0}}\sigma _{A,total}\label{mathed:f''A(sigma)}\end{equation}
 where $\sigma _{A,total}(E)=\sigma _{elastic\, scatt.}+\sigma _{abs.}$
and $r_{0}=\frac{q^{2}}{4\pi \varepsilon _{0}mc^{2}}$ is the classic
electron radius. The absorption coefficient $\mu _{A}(E)=N_{A}\sigma _{A,total}(E)$,
where $N_{A}$ is the number of atoms \emph{A} per volume unit, is
the quantity measured by XAFS spectroscopy. Causality implies that
$f_{A}^{\prime }$ (dispersion) and $f_{A}^{\prime \prime }$ (absorption)
are not independent, but rather related by the Kramers-Kronig transforms,
as shown in figure \ref{fig:fpsfs_InAs}. The virtual, intermediate
state of the resonant scattering process corresponds to an atom with
a core hole and a virtual photoelectron in an atomic-like electronic
state or in the continuum. Similar to x-ray absorption, the energy
dependence above the Fermi level of the electronic density is shaped
and modulated by the local atomic environment of the resonant atom.
The factors $f_{A}^{\prime }$ and $f_{A}^{\prime \prime }$ take
into account the transition to intermediate states ; for convenience,
in the extended region, they can be split in {}``smooth'' and oscillatory
parts : $f_{Aj}^{\prime }=f_{0A}^{\prime }+\Delta f_{0A}^{\prime \prime }\chi _{Aj}^{\prime }$
and $f_{Aj}^{\prime \prime }=f_{0A}^{\prime \prime }+\Delta f_{0A}^{\prime \prime }\chi _{Aj}^{\prime \prime }$,
where $f_{0A}^{\prime }$ and $f_{0A}^{\prime \prime }$ are the resonant
scattering terms of an isolated atom \emph{A}, $\Delta f_{0A}^{\prime \prime }$
represents the contribution of the resonant scattering to $f_{0A}^{\prime \prime }$
; $\chi _{Aj}^{\prime }$ and $\chi _{Aj}^{\prime \prime }$ are the
fine structure oscillations\emph{.}

\subsection{Selectivity}

Unlike XAFS, DAFS spectroscopy is site-selective because it measures
the x-ray scattering and thus takes advantage of the wave interference
to select a subset of atoms in the sample. For the sake of clarity,
consider a centrosymmetric, periodic structure (the general case is
considered in references \cite{Meyer98,Proietti99}). The scattering
structure factor is \begin{equation}
F\propto \sum _{j}m_{j}c_{j}\exp (-M_{j}Q^{2})f_{j}(\vec{Q},E)\cos (\vec{Q}.\vec{r_{j}})\label{eq:facteur_structure}\end{equation}
where the vector $\vec{r}_{j}$ gives the atomic position in the cell,
$m_{j}$, $c_{j}$, $\exp (-M_{j}Q^{2})$ are the multiplicity, occupation
and Debye-Waller factors respectively. The scattered intensity is
then proportional to the squared modulus of the structure factor.
The summation runs over all atoms in the unit cell that are not related
through the center of symmetry. The site-selectivity is given by the
term $\cos (\overrightarrow{Q}.\vec{r}_{j})$, the value of which
depends on the dot-product of the momentum transfer vector with the
position of atom \emph{j} in the cell. Because of this, DAFS may distinguish
the valence and local environment of atoms of the same atomic number
located on different crystallographic sites. This has been used, for
example, to recover the Fe scattering anomalous terms of the tetrahedral
and octahedral sites in magnetite \cite{Pickering93b}, the local
environment of Cu1 and Cu2 sites in the superconductor $YBa_{2}Cu_{3}O_{7}$
\cite{Cross97a,Cross98b}, and the interface structure of an Ir/Fe
superlattice \cite{Renevier97prl}. In the special case of magnetite,
there exist Bragg reflections strictly sensitive either to the octahedral
or tetrahedral sites and without contribution from the other site.
To apply site-selectivity and recover site-selective oscillations
the long range ordered average crystallographic structure must be
known precisely. On the other hand, it is well known that Multi-wavelength
Anomalous Diffraction (MAD) gives valuable information to solve the
crystallographic structure (see for instance \cite{Hendrickson85}).
Another useful application of DAFS is \emph{spatial-selectivity}.
Information about a material embedded in another one or which coexists
with others can be obtained by measuring the DAFS spectra of the Bragg
peaks of that material. Typical examples are thin films having one
or more atomic species in common with the substrate and/or buffer
(see examples in reference \cite{Hodeau2001}). More recently, spatially-selective
DAFS has been applied to nanostructures \cite{Grenier02} ; in these
cases, XAFS experiment in fluorescence mode would not allow discrimination
of signals from heterostructures with a common element.

\subsection{Polarization dependence}

As for XAFS, DAFS spectra depend on the x-ray beam polarization direction.
For x-ray absorption, the crystallographic point group governs that
dependence. For instance the absorption is isotropic for a cubic point
group even if the site symmetry of the absorber is not \cite{Brouder90}.
In case of DAFS instead, the situation is quite different because
of the interference mechanism. The polarization of the incoming and
outgoing x-ray beams must be taken into account. Via virtual mutipole
transitions (mainly dipolar, quadrupolar or an interference of both),
the energy dependences in $\sigma -\sigma $ and $\sigma -\pi $ channels
(see section \ref{sec:data_reduction} and figure \ref{fig:scatt_geom_5.28})
upon the incoming x-ray beam energy as well as upon the azimuthal
angle (corresponding to a rotation about the scattering vector) can
reveal the Anisotropy of the Susceptibility Tensor (ATS) \cite{Templeton80,Dmitrienko83}.
The anisotropy is due to, for instance, d-orbital ordering \cite{Murakami98},
distortion ordering \cite{Garcia00}, charge ordering \cite{GrenierS01}.
Regarding structural analysis, the polarization dependence may be
used as in x-ray absorption. For instance, in strained thin films
or superlattices of materials which are cubic in the bulk, the interfaces
may be non cubic at the atomic scale. In that case, it is worth measuring
DAFS spectra with the polarization of the incoming beam in and out
of the growth plane to probe in- and out of plane local distortions
\cite{Grenier02}.

\section{Energy-Scan Diffraction}

The DAFS experiment is time consuming since intensity oscillations
of several Bragg reflections must be collected as a function of energy
over a few hundreds of electron-volts (eV) with a typical energy resolution
of 1 eV, and sometimes require measurement of tiny diffraction intensity
variations. This is the reason why the techniques for experimental
set-up and data collection are still under development. We report
on the progress of the Energy-Scan Diffraction (ESD) with optics for
delivering monochromatic beam at the beam line BM2-D2AM at the ESRF.
In this mode, the intensity of a Bragg reflection is collected as
a function of the energy given by the Bragg angle of the monochromator
crystal and is tuned step by step through the absorption edge of one
of the atomic species in the sample. The requirements of the experiment
are to measure the intensities versus the energy as fast as possible,
with a monitor corrected signal-to-noise ratio as high as 1000 or
more (comparable to a typical XAFS experiment), and without distortions
of the spectra. These requirements turn out to be a challenge for
diffraction experiments. For that purpose, we use a fixed-exit diffraction
beamline with a XAFS monochromator coupled to a diffractometer both
having high precision movement (<1/1000°). The beamline BM2-D2AM was
built for using the light emitted by a bending magnet and is dedicated
to anomalous scattering experiments \cite{Ferrer98}. The beamline
optics consist of a double-crystal, channel-cut geometry, water cooled
monochromator and two 1.2 meter long platinium-coated Si-mirrors (a
schematic representation in given in reference \cite{Ferrer98}, figures
1 and 6). The two mirrors are used for harmonic rejection, for vertical
focusing, and for vertical stability of the incoming x-ray beam at
the sample position when changing the x-ray energy. The first mirror,
which is upstream of the monochromator, also removes part of the heat
load. The second crystal of the monochromator, usually Si(111), is
mounted onto a bender \cite{Hazemann95} providing dynamic sagittal
focusing. The lowest spot size at the focal point that can be delivered
with the optics is 100$\mu $m horizontal by 100$\mu $m vertical.
A 7-circle diffractometer allows diffraction measurement in the vertical
and horizontal planes and is suitable for polarization-dependent DAFS
experiments. A crystal analyser, a Displex cryostat, a furnace, and
spherical beryllium vacuum enclosures are also available. It is worth
emphasizing that reliable DAFS spectra are obtained provided that
the beam size and position at the sample position (usually the focal
point) are very stable over the energy range covered by the energy
scan (about 1keV). This implies perfect alignment of the optics, including
dynamic sagittal focusing and tuning of the second crystal of the
mochromator.

\subsection{Maximum intensity measurements}

\label{sec:max_intensity_measurement} 
\begin{figure}[htbp]
\begin{center}\subfigure[]{\includegraphics[  width=0.49\textwidth,
  keepaspectratio]{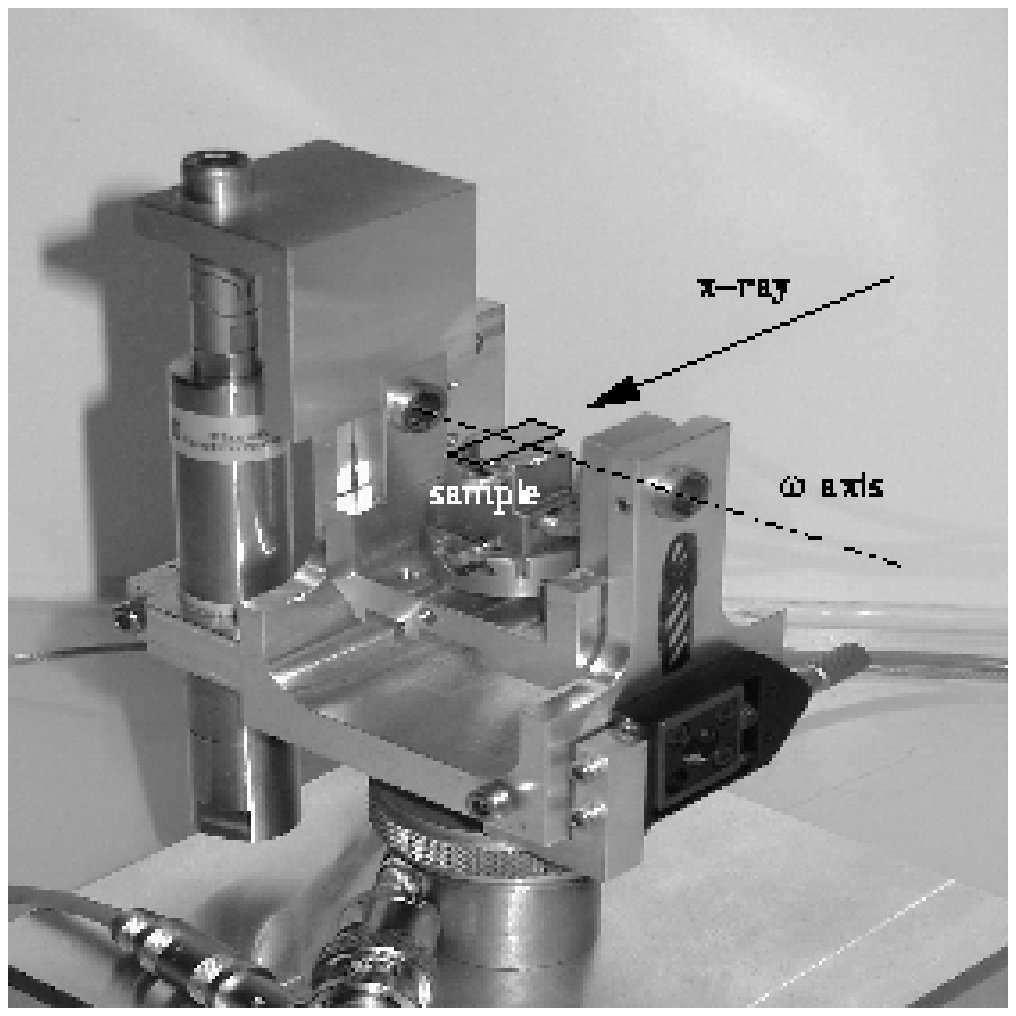}}~~\subfigure[]{\includegraphics[  width=0.49\textwidth,
  keepaspectratio]{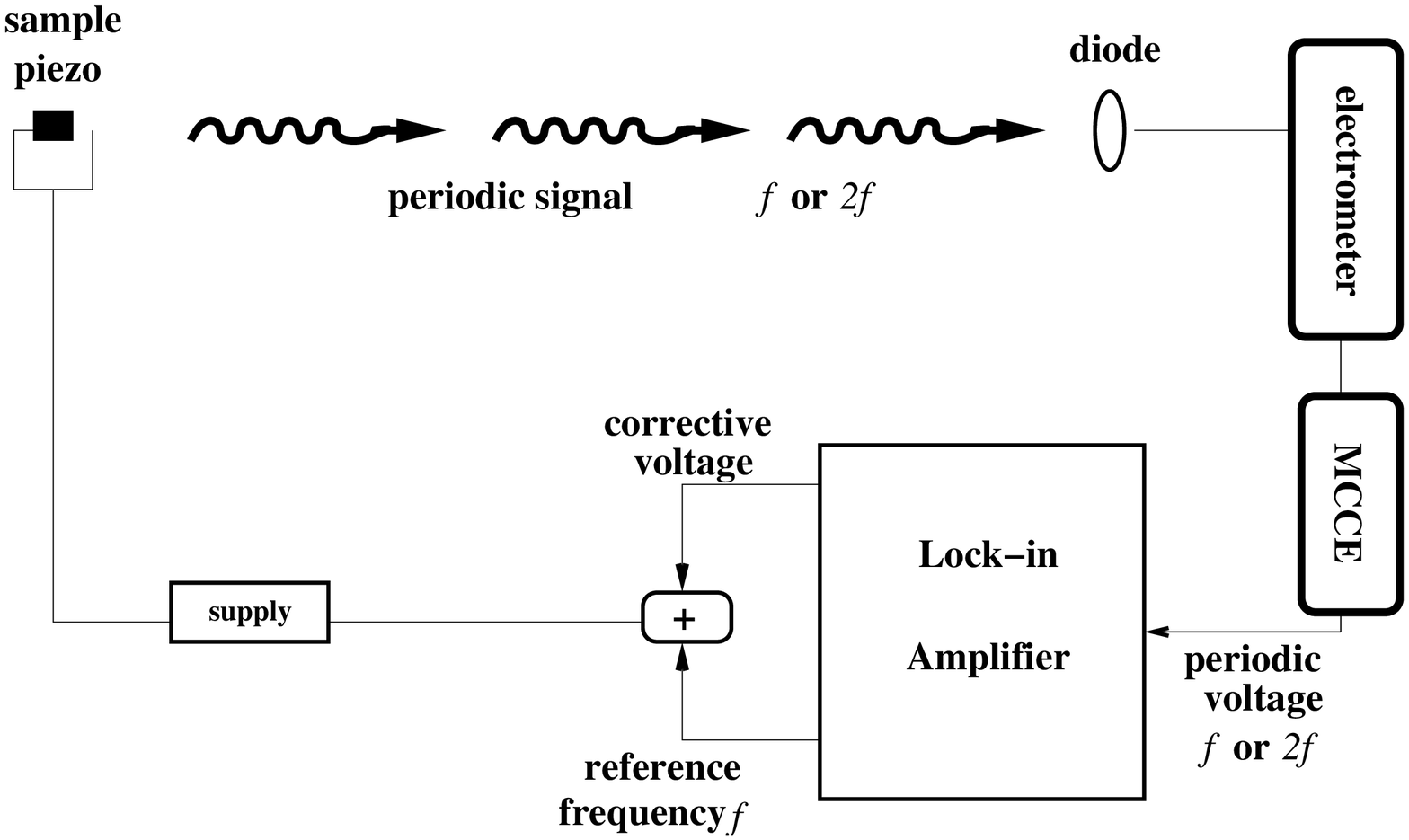}}\end{center}

\caption{\label{fig:rocking-sample-holder 3.5} (a) A picture of the rocking
sample holder developed at beamline BM2-D2AM. One can see the shiny
surface of a \emph{GaAsP} sample on the goniometer head. (b) A schematic
of the feedback set-up.}
\end{figure}

Because \emph{integrating}
reflection profiles at each energy and at several diffraction vector
is time consuming, much effort have been made to perform the data
collection by measuring only the maximum intensity of the Bragg peak
as a function of energy. This allows minimizing the contribution of
the fluorescence signal of the sample and so maximizing the fraction
of diffracted photons measured in the diffraction detector. The measurement
of the maximum of the diffracted intensity gives results identical
to the integration measurement provided that the rocking-curve reflection-profile
is regular, as with thin films, superlattices, and heterostructures.
With bulk material or powder, a drastic change of the absorption length
may lead to profile variations above the absorption edge. The peak
profile depends on the mosaïcity, the vertical beam divergence, and
the beam coherence. In that sense, the high beam collimation and coherence
produced by third generation insertion-device sources, may be a concern.
Since the mechanical precison and stability of all motors is necessarily
limited, we developed a feedback control of the sample rocking-angle
position (based on a sample-holder that rocks the sample around the
\emph{omega} axis, see figure \ref{fig:rocking-sample-holder 3.5}a),
for measuring the maximum intensity \cite{Cross98jsr,Blanco98,Grenier}.
The idea is to measure a quantity that is proportional to the derivative
of the Bragg peak profile. At the low and high angle sides of the
peak maximum the derivative is positive and negative respectively,
whereas at the maximum is equal to zero. The derivative is then used
to correct the \emph{omega}-angle in order to have the maximum intensity
in the diffraction detector. When the derivative is equal to zero
the sample angle position is stable. Figure \ref{fig:rocking-sample-holder 3.5}b
shows the scheme of the set-up developed at BM2-D2AM. The diffraction
signal $I_{d}$ is modulated by a reference sine signal rocking the
sample holder and delivered by the lock-in amplifier (EG\&G instruments,
7220 DSP) and fed in the lock-in input. The output is then given by
$I_{out}=\frac{1}{T}I_{d}[\omega _{0}+\Delta \omega \sin (\Omega t+\varphi )]\sin (\Omega t)dt$,
where $T=\frac{2\pi }{\Omega }$ is the period of the oscillation,
$\varphi $ the phase shift between $I_{d}$ and the reference signal
that is set to 180°. For amplitude modulations smaller than the Full
Width at Half Maximum (FWHM), it can be easily shown with a Taylor
expansion of the diffracted intensity to first order that \begin{equation}
I_{out}\simeq \frac{1}{T}\int I_{d}(\omega _{0})\sin (\Omega t+\varphi )dt+\frac{1}{T}\int [\frac{dI_{d}}{d\omega }\Delta \omega \sin (\Omega t+\varphi )]\sin (\Omega t)dt\label{mathed:I_out}\end{equation}
The first term is equal to zero, whereas the second is proportional
to $\frac{dI_{d}}{d\omega }\Delta \omega \cos \varphi $. A piezoelectric
transductor (PI P-843.40) produces oscillations with amplitudes $\Delta \omega $=0.0167°/V
and frequencies in the range of 10-20Hz. By compensating the mechanical
imperfections, the feedback control ensures measuring the maximum
intensity of the \emph{omega}-profile as a function of the energy.
It leads to a significant improvement of the signal-to-noise ratio
and reduction of the low-frequency distortion of the spectra. Future
improvements of this device include rocking the sample in arbitrary
orientation and its implementation in high and low temperatures sample
environments. The feedback has broadened the application of the maximum
intensity measurement procedure to very narrow \emph{omega}-profiles\emph{,
i.e.} with a FWHM of the order or less than 0.01° in an \emph{omega}-scan.

\subsection{Integration of the Bragg peak profile }

In the case of mosaïcity and/or domains, which introduce irregular
reflection shapes \cite{Vacinova95a}, or strong absorption \cite{Bos}
that modify the Bragg peak profile when crossing the absorption edge,
an \emph{omega} scan (perpendicular to the diffraction vector) or
an \emph{omega}-\emph{two}-\emph{theta} scan (along the diffraction
vector) is mandatory to integrate the peak profile. To reduce the
measurement time, a flying \emph{omega}-scan is performed at each
energy step together with a counting on both sides of the peak to
measure the background, including the sample fluorescence. In practice,
the counting is triggered by the \emph{omega} motor and the integrated
intensity is the total number read by the counting crate during the
time elapsed by the motor movement. This procedure is time consuming
and the ratio of diffracted intensity to background signal is inherently
much smaller compared to the maximum intensity mesurement.

\subsection{Experimental set-up and procedure}

\begin{figure}[htbp]
\begin{center}\includegraphics[  width=0.70\textwidth,
  keepaspectratio]{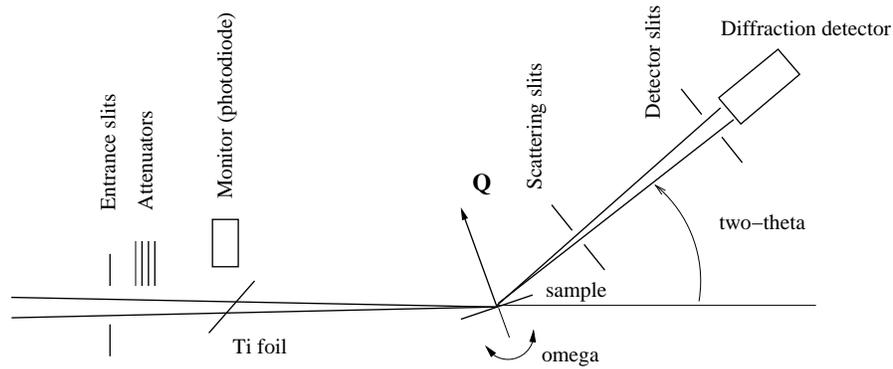}\end{center}

\caption{\label{cap:diffraction-set-up} Schematic of the in-hutch portion
of the diffraction set-up used to measure DAFS. Detailed information
on detectors is given in section \ref{sub:Detectors}.}
\end{figure}

Our purpose is to obtain DAFS with data quality comparable to typical
XAFS data quality, thus making DAFS attractive to the XAFS community.
This means that the diffraction is to be measured with a high signal-to-noise
ratio as a function of the energy (at least 1000) and without any
distortions. The experimental diffraction set-up (figure \ref{cap:diffraction-set-up})
consists of entrance slits, attenuators, and a monitor (\emph{I$_{0}$}),
all in vacuum environment. The sample, mounted on a 7-circle diffractometer,
the scattering slits and the detector slits all lie outside the vacuum.
As for a XAFS experiment, the incoming beam must be carefully monitored.
This is done by measuring the fluorescence signal emitted by a 4$\mu $m
thick, \emph{}99.6\% \emph{}pure \emph{Ti} foil mounted in vacuum
and at 45° with respect to the beam path. Homogeneity of the foil
and high counting rate of the fluorescence signal ensure a high signal-to-noise
ratio. It is important to check that the normalization of the diffracted
signal by \emph{I$_{0}$} suppresses any fluctuations in the incoming
x-ray beam as a function of time and energy. Much effort must be paid
to the optical alignment and dynamic sagittal focusing and tuning
of the second crystal of the monochromator are needed in order to
assure a smooth \emph{I$_{0}$ signal} and beam stability as a function
of energy. Also, one must check that the beam is well aligned through
the centers of all sets of slits. All attenuators are to be located
beyond the monitor so as not to perturb \emph{I$_{0}$} when they
are inserted in the beam. However, it is preferable not to use attenuators
for measuring a DAFS spectrum. Regarding the detector and scattering
slits, once the optimization of the Bragg diffraction has been achieved,
a balance is struck between opening the slits enough to accept the
divergence of the diffracted beam (with attention paid to the loss
of resolution in the reciprocal space) and minimizing the fluorescence
signal of the sample. High signal-to-noise ratio is most easily obtained
with flat samples which intercept the whole-diffracted beam. Additionally,
flat samples are easier to correct for the effect of absorption, as
described in section \ref{sub:Absorption-correction}.

\begin{figure}[htbp]
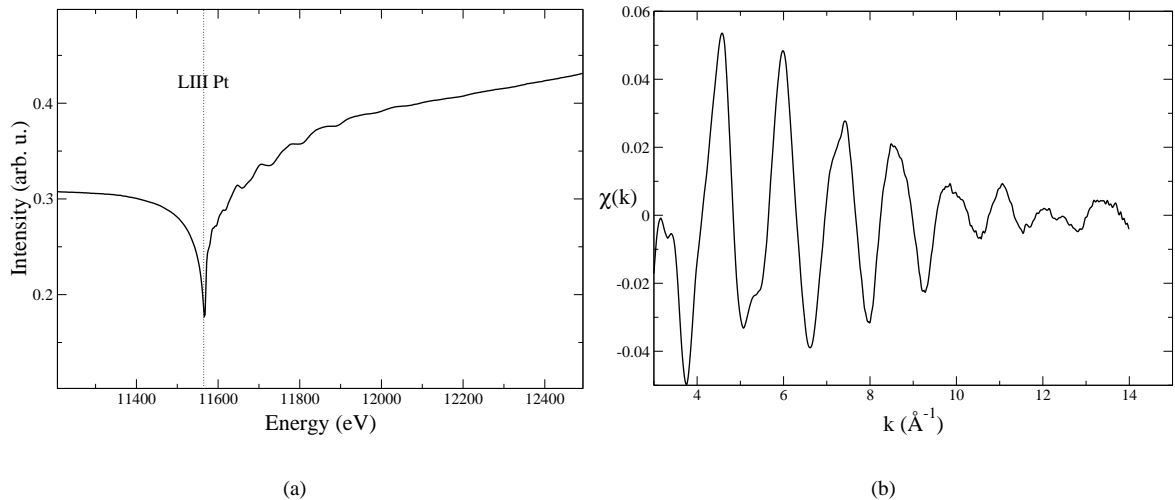

\vspace*{\medskipamount}
\begin{center}\subfigure[]{\includegraphics[  width=0.45\textwidth]{fig4a.eps}}~~\subfigure[]{\includegraphics[  width=0.45\textwidth,
  keepaspectratio]{fig4b.eps}}\end{center}

\caption{\label{fig best_dafs+edafs} (a) Raw DAFS spectrum $I_{d}/I_{0}$
of the (00l) superstructure reflection of a 50nm thick CoPt thin film,
measured around the Pt $L_{III}$ edge. (b) The EDAFS (Extended DAFS)
spectrum normalized to $I_{s}$, where $I_{s}$ is the diffracted
intensity without oscillations.}
\end{figure}

For measuring a DAFS spectrum, either by measuring the maximum intensity
or by integrating through the peak profile, one needs to track the
Bragg peak as a function of the energy. For this, the intensity of
the Bragg peak is carefully optimized at 3 or 4 points in the energy
range of interest. Then a linear regression of sample position to
energy is made to calculate the \emph{omega} and \emph{two}-\emph{theta}
angles throughout the energy scan. Note that the backlashes of all
motors involved during a scan must be carefully checked and accounted
for. As for a XAFS experiments, the linearity versus the energy of
both detectors measuring the incoming and diffracted beams must be
checked. As a example of typical data obtained with thin films or
superlattices, figure \ref{fig best_dafs+edafs} shows the DAFS spectra
of a (001) superstructure reflection of a 50nm thick CoPt thin film,
measured at the Pt $L_{III}$ edge \cite{Ersen01}. 

\subsection{Background subtraction.}

The energy-dependent background due to the fluorescence of the sample
\textbf{}may not be negligible in the detector \textbf{}when measuring
weak reflections. The use of an energy discriminating detectors to
suppress the fluorescence is often not possible due to their very
low counting rates. A simple solution is to collect a spectrum off
the Bragg peak and \textbf{}then to subtract it from the DAFS spectrum
measured at the maximum intensity. However, a less time-consuming
method is desirable. \textbf{}Wavelength discriminating detection
with a crystal analyzer is also used. For reasons of efficiency, an
analyzer should have an angular acceptance comparable to the divergence
of the diffracted beam and a reflectivity as high as possible. On
this purpose, we use the (002) reflection of a flat graphite single
crystal, the mosaïcity of which is about 0.3° (FWHM). When mesuring
very low counting rates, care should be taken of all fluorescence
emission lines, as for instance the $K_{\beta }$ that can not be
isolated from the elastic signal at the edge. It should be noted that
the use of an analyzer crystal adds at least one extra motor movement
to adapt the crystal angle as the x-ray beam energy is tuned. We also
have carried out experiments at BM2-D2AM using the (222) reflection
of an MgO crystal. Although the beam attenuation was about 40 in comparison
with the intensity measured without crystal, we were able to measure
DAFS spectra of the weak, forbidden reflections (002) and (006) of
a $Fe_{3}O_{4}$ single crystal at the Fe K-edge \cite{Garcia00}.
Graded lattice spacing multilayers are good candidates because of
their large angular acceptance and high reflectivity. A further alternative
is dynamic background subtraction. As described above in section \ref{sec:max_intensity_measurement},
it is possible to measure the derivative of the Bragg peak profile,
therefore the constant background, including the fluorescence signal
of the sample, is readily eliminated. A double integration of the
derivative signal gives back the integrated intensity over the background
\cite{Coraux00}.

\subsection{Detectors}

\label{sub:Detectors}

\begin{figure}[htbp]
\begin{center}\includegraphics[  width=0.49\textwidth,
  keepaspectratio]{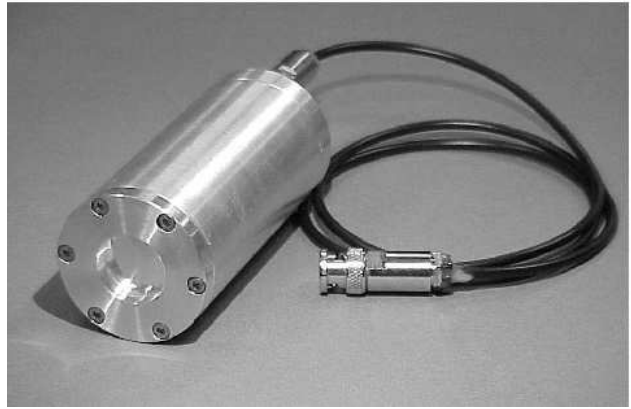}\end{center}

\caption{\label{fig picture of photodiode} A photodiode-based detector used
for measuring DAFS spectra. The dimensions are the same as for common
scintillator-based detectors, allowing the use of off-the-shelf detector
mounts.}
\end{figure}

To cope with and to benefit from the high counting rate on the $I_{0}$-detector,
the sample fluorescence, or the diffraction peak, we have developed
a detector based on photodiodes operating in photovoltaic mode at
room temperature which provides high linearity and very large dynamic
range. On the other hand, photodiodes have a high Quantum Efficiency
(about 4eV is necessary to create an electron-hole pair in Si) and
a very low background can be achieved, allowing to count x-ray photons
down to few thousand per second at 10keV. These are PIN Silicon photodiodes
(Canberra-Eurysis, PD 300-15-300 CB) with active area diameter of
19.5 mm, 300$\mu $m thick, and covered with a 50m thick aluminium
entrance window. The surface orientation is {[}111{]} with a miscut
of 7°. Figure \ref{fig picture of photodiode} shows our photodiode-based
detector. For practical convenience the external design is the same
as a scintillator detector. There is no metal or ceramic support right
back the photodiodes and the connection on the silicium is shielded.
A 10 micron thick aluminum foil prevents illumination of the photodiode
by visible light. High purity aluminum foil (99.999\%) is used to
avoid signal contamination by trace elements such as iron or iron-oxide,
although high purity Beryllium windows would be an attractive alternative
for low-energy DAFS. To connect the photodiodes to the electrometer,
we use low-noise tri-ax cables (Keithley SC22) as short as possible
to limit current fluctuations (the maximum length is 1 meter). Consequently,
the electrometer for measuring the diffracted beam is mounted on the
\emph{two-theta} arm. The photo-current is measured with a NOVELEC$^{\textrm{TM}}$
electrometer (EC-PV High Sensitivity) with current scales ranging
from $10^{-10}$ to $3\times 10^{-7}$A. The V/F converter starts
at $10^{6}$ Hz (up to $10^{7}$ Hz) to ensure linearity of the V/F
conversion at very low voltage, allowing a precise compensation of
the leakage current. The leakage current is lower than $10^{-12}$
A and can barely be detected. Most important, the dark current signal-to-noise
ratio is about $10^{4}$. However, due to temperature variations in
the experimental hutch, the dark current can fluctuate with an amplitude
lower than $\pm 5\times 10^{-13}$A on a time scale of few hours.
Therefore a periodic measurement of the dark current is performed
during the data acquisition when measuring with the most sensitive
range ($10^{-10}$A). A further development will be inserting a chopper
in the beam (with a frequency of few tenth of Hertz) and performing
a synchronous detection to remove the dark current .

These detectors have a very large dynamic range, starting from a thousand
counts per second (cps) up to the actual upper limit given by the
highest scale ($3\times 10^{-7}$A), corresponding to about $9\times 10^{8}$
cps at 10keV. Although the photodiodes are well suited to a typical
DAFS experiment, multiple diffraction inside the photodiode might
be excited. This would be true in a case where the diffracted beam
has a very small divergence, comparable to the that of the perfect
Si crystal.

\section{Quick DAFS}

\begin{figure}[htbp]
\begin{center}\subfigure[]{\includegraphics[  width=0.49\textwidth,
  keepaspectratio]{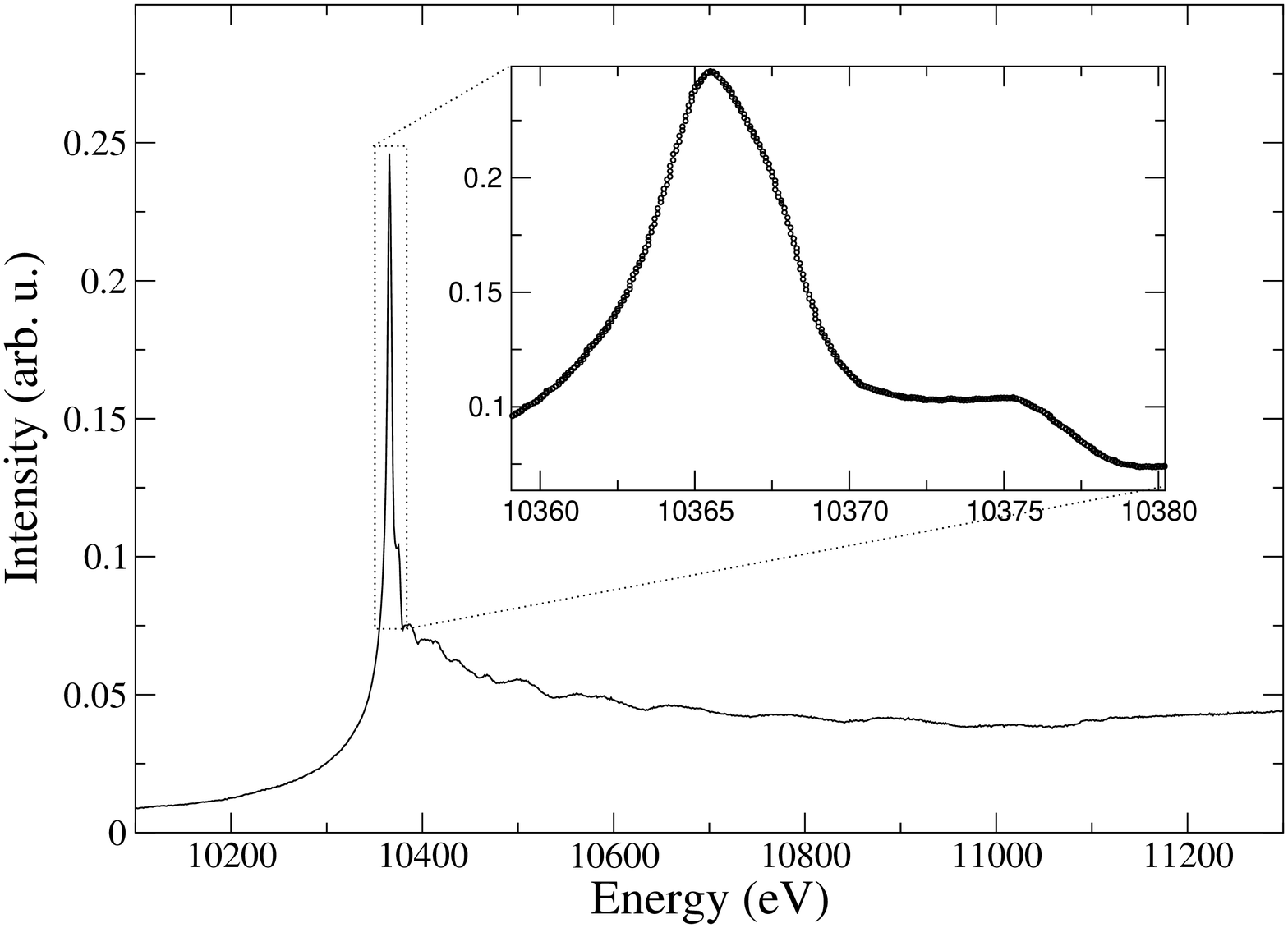}}\subfigure[]{\includegraphics[  width=0.49\textwidth]{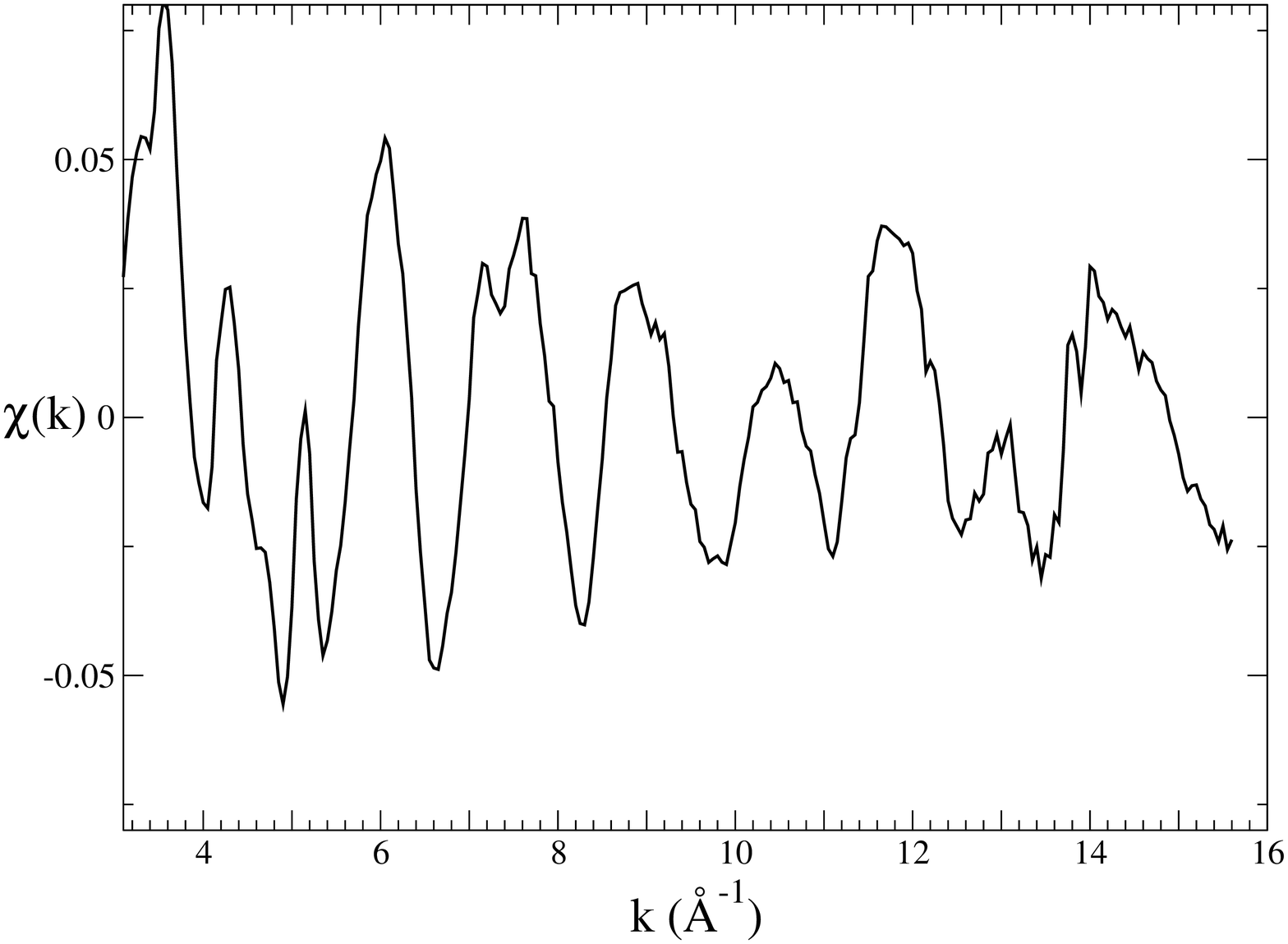}}\end{center}

\begin{center}\subfigure[]{\includegraphics[  width=0.49\textwidth,
  keepaspectratio]{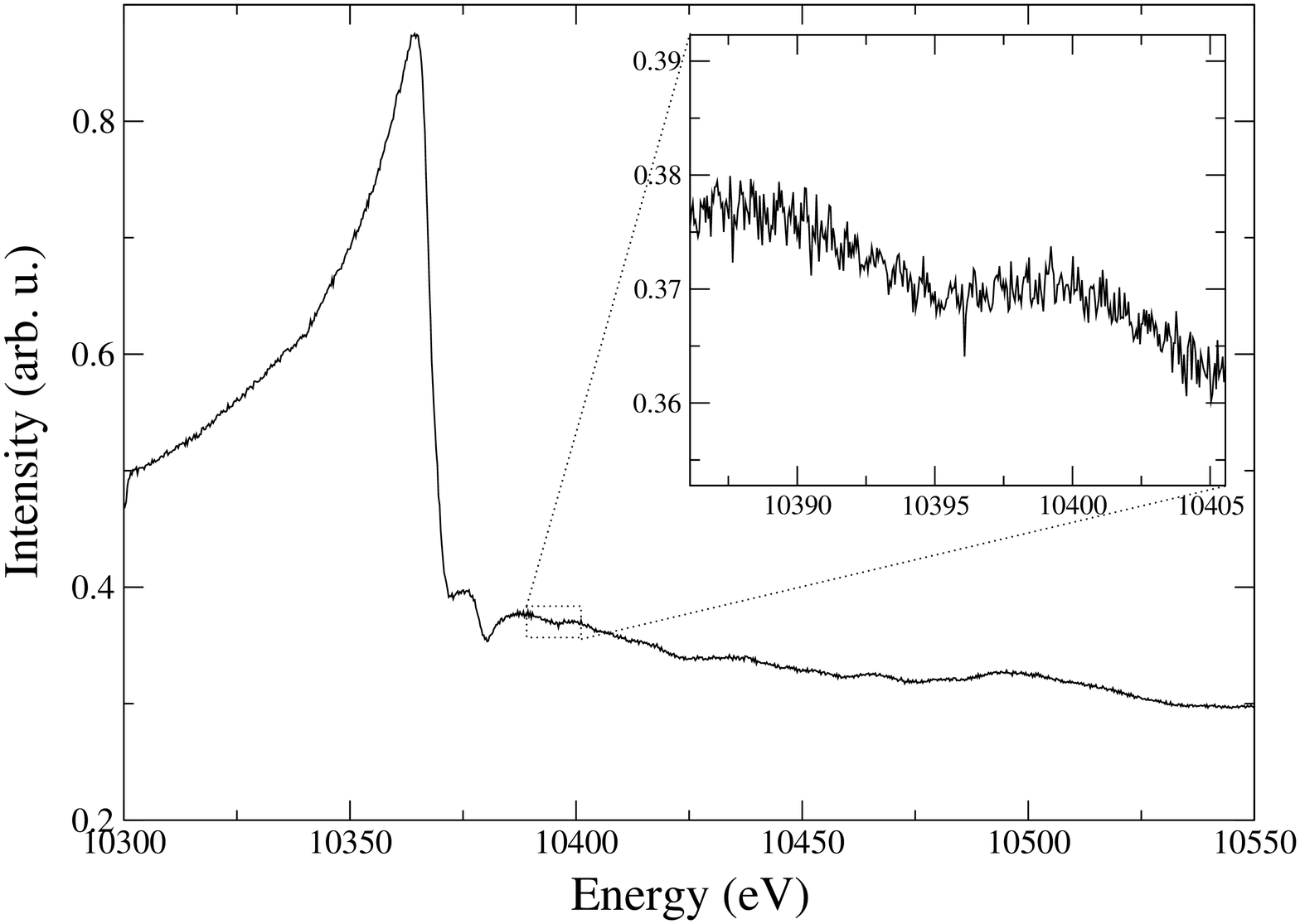}}\end{center}

\caption{\label{fig q-DAFS assist=E9 3.13a & 3.16 & 3.17b} (a) q-DAFS spectrum
of the (006) reflection of a GaAsP film, measured in 4 minutes using
the feedback control described in section \ref{sec:max_intensity_measurement}.
(b) EDAFS (Extended-DAFS) of the (006) reflection of the GaAsP file,
normalized to $I_{s}$. (c) q-DAFS spectrum of the (006) reflection
of a GaAs substrate, with the feedback control. The FWHM of the substrate
peak was 0.008°. Insets show the signal quality in terms of sampling
(0.05eV step) and signal-to noise-ratio.}
\end{figure}

Currently, data collection time for an Energy-Scanned experiment is
rather long, on the order of one or a few hours. This is mainly due
to the dead time required to move the sample and monochromator motors
and is not due to the counting time which is often as short as a few
seconds. To address this, we have implemented a quick-DAFS (q-DAFS)
procedure that allows counting during motor movements. Quick-EXAFS
has been used for more than a decade, since the pioneering work of
Frahm et al. \cite{Frahm88}, and a one keV range quick-EXAFS spectrum
is typically obtained in a few seconds. A q-DAFS experiment is more
complicated since one need to move at least three motors: the monochromator,
\emph{omega} and \emph{two}-\emph{theta}. Here we show how we have
achieved a one keV range q-DAFS spectrum in a few minutes, as compared
to about one hour for a step-by-step scan. The principle is to use
the pulses generated by the monochromator stepper motor to trigger
the stepper motion of the others motors as well as the counting crate
in order to have perfect synchronization. The angular speed of the
monochromator ($\frac{d\omega _{mono.}}{dt}$) is related to the \emph{omega}
motor speed of the diffractometer ($\frac{d\omega }{dt}$) by:

\begin{equation}
\frac{d\omega _{mono.}}{dt}=\frac{d\omega }{dt}\, \sqrt{\frac{(\Gamma _{mono.}E)^{2}-1}{(\Gamma _{sample}E)^{2}-1}}\label{mathed:q_dafs}\end{equation}
where $\Gamma =\frac{2d_{hkl}}{hc}$ and \emph{E} is the energy of
the incoming beam ($\Gamma =\frac{2d_{hkl}(\textrm{Å})}{12.398}$).
Equation \ref{mathed:q_dafs} shows that the angular speeds are related
via a non-linear expression, except when $\Gamma _{sample}E\gg 1$
and $\Gamma _{mono.}E\gg 1$, for example at high energy. This difficulty
is solved with the programmable VPAP (VME-PAP : VERSA Module Eurocard-PAP)
crates that drive the stepper motors and allow arbitrary synchronization
of the motor movements. The monochromator motor is launched with a
given angular speed from the initial to final $\omega _{mono.}$ values.
The \emph{omega} and \emph{two}-\emph{theta} motors move, one and
two step respectively, in the forward direction according to Bragg's
law when a tabulated number of pulses from the \emph{}monochromator
motor is intercepted by the VPAP crates. We also synchronize the movement
of a few other motors as a function of the energy to dynamically focus
the beam in the horizontal plane and to tune the monochromator second
crystal. Every pulse from the monochromator motor triggers the reading
of the counting crate. The actual minimum angular step of that motor
is 1/20000°. With a monochromator angular speed of 100 steps/s, a
reading is done every 10ms and about every 0.05eV at the Ga K-edge
(10.367 keV). This means that 200s are needed to measure a DAFS spectrum
over a 1000$\, $eV range. This hardware based procedure ensures a
perfect synchronism of all motors and counting and can also be used
with the feedback control described in section \ref{sec:max_intensity_measurement}.
The angular speed of the monochromator depends on the number of diffracted
photons per second, and should be adapted accordingly to obtain an
appropriate counting statistic. It also depends on the time-response
and efficiency of the sample rocking-angle feedback. We have tested
the q-DAFS with a GaAsP thin film grown on a GaAs substrate. Figure
\ref{fig q-DAFS assist=E9 3.13a & 3.16 & 3.17b}a shows the DAFS spectrum
of the (006) reflection of the GaAsP film, measured in 4 minutes,
with the feedback control on, and figure \ref{fig q-DAFS assist=E9 3.13a & 3.16 & 3.17b}b
shows the Extended-DAFS extracted using the program \textsc{autobk}
\cite{Autobk93} and the formula $\frac{I_{meas.}-I_{s}}{I_{s}}$,
where $I_{s}$ corresponds to the diffracted intensity without oscillations.
The full width at half maximum of the rocking scan was 0.02°. Similar
spectra, reported in reference \cite{Proietti99}, were obtained using
the standard step-by-step energy scan and by measuring the maximum
intensity, without feedback, in about 1 hour per scan. Indeed, the
q-DAFS procedure gives spectra of similar quality in much shorter
time measurement. With the feed back control it was also possible
to measure the q-DAFS spectrum of the very narrow (006) reflection
(FWHM=0.008°) of the GaAs substrate, shown in see figure \ref{fig q-DAFS assist=E9 3.13a & 3.16 & 3.17b}c.

\section{Data reduction for EDAFS analysis}

\subsection{Diffraction intensity}

\label{sec:data_reduction}For any crystallographic structure, the
structure factor may be written \cite{Proietti99} :

\begin{equation}
F(\overrightarrow{Q},E)=\left\Vert F_{T}(\overrightarrow{Q},E)\right\Vert e^{i\varphi _{T}(\vec{Q})}+\sum _{j=1}^{N_{A}}\alpha _{Aj}(\overrightarrow{Q})e^{i\varphi _{Aj}(\vec{Q})}\left(f_{Aj}^{\prime }+if_{Aj}^{\prime \prime }\right)\label{mathed:F(Q,E)}\end{equation}

\begin{figure}[htbp]
\begin{center}\includegraphics[  width=0.40\textwidth]{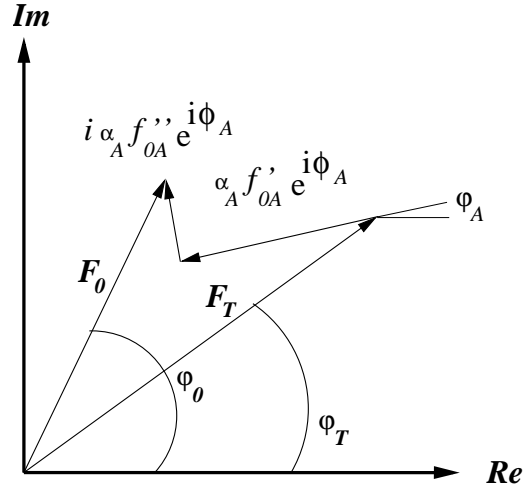}\end{center}

\caption{\label{fig:F0_decomposition} Representation of the structure factor
$F_{0}(\vec{Q},E)$ in the complex plane (see text). }
\end{figure}

The summation runs over all anomalous atoms in the unit cell, $N_{A}$,
$\vec{Q}$ is the scattering vector and $E$ the energy of the incident
beam. $F_{T}$ is a complex structure factor of phase $\varphi _{T}$
which includes the overall contribution of all non anomalous atoms
and the Thomson scattering of all anomalous atoms, $\alpha _{Aj}=c_{Aj}\exp (-M_{Aj}Q^{2})$,
$\varphi _{Aj}=\vec{Q}.\overrightarrow{r}_{Aj}$ the scattering phase
of atom \emph{A} on site \emph{j}. Then it can easily be shown that
: 

\begin{equation}
\left\Vert F_{0}(\vec{Q},E)\right\Vert ^{2}=\left\Vert F_{T}(\vec{Q},E)\right\Vert ^{2}\left[\left(\cos (\varphi _{T}-\varphi _{A})+\beta f_{0A}^{\prime }\right)^{2}+\left(\sin (\varphi _{T}-\varphi _{A})+\beta f_{0A}^{\prime \prime }\right)^{2}\right]\label{mathed:F0_square}\end{equation}
where, $F_{0}(\vec{Q},E)$ is the smooth, non-oscillatory part of
the complex structure factor and $\varphi _{0}(\vec{Q},E)$ its phase,
calculated assuming that the anomalous corrections ($f_{0A}^{\prime }$,$f_{0A}^{\prime \prime }$)
are identical for all anomalous atoms in the EDAFS region, $\beta =\frac{\left\Vert \alpha _{A}\right\Vert }{\left\Vert F_{T}\right\Vert }$,
$\left\Vert \alpha _{A}\right\Vert e^{i\varphi _{A}}=\sum _{j=1}^{N_{A}}\alpha _{Aj}e^{i\varphi _{Aj}}$.
The decomposition of $F_{0}(\vec{Q},E)=\left\Vert F_{T}(\overrightarrow{Q},E)\right\Vert e^{i\varphi _{T}(\vec{Q})}+\left\Vert \alpha _{A}\right\Vert e^{i\varphi _{A}}\left(f_{0A}^{\prime }+if_{0A}^{\prime \prime }\right)$
in the complex plane is shown in figure \ref{fig:F0_decomposition}.
Equation \ref{mathed:F0_square} shows that the energy dependent variations
of the diffracted intensity near an absorption edge give access to
the phase difference $\Delta \varphi =\varphi _{T}-\varphi _{A}$
and the ratio $\beta $, and therefore give precise information on
the long-range average crystallographic structure, i.e. on atomic
displacements and concentration of the anomalous atoms. The smooth
part of a DAFS spectrum can be simulated by equation \ref{mathed:F0_square},
although there may be some discrepancy at the edge for weak reflections
if different anomalous sites exist with energy shifted (also known
as chemically shifted) anomalous factors. Apart from recovering the
crystallographic parameters $\Delta \varphi $ and $\beta $, equation
\ref{mathed:F0_square} is used to verify the absence of gross data
distorsions throughout the energy range. Equations \ref{mathed:F(Q,E)}
and \ref{mathed:F0_square}, can be easily generalized to the case
of two or more chemically shifted resonant atoms or to the case of
atoms with different atomic numbers but nearby resonant energies \cite{Ravel99prb}.

A DAFS spectrum is related to the square modulus of the structure
factor after several corrections and according to the following formula
:\begin{equation}
I_{bs}=I_{meas}-I_{bgd}=r_{0}^{2}S(\vec{Q})A(\vec{Q},E)D(E)\frac{1}{sin\theta }L(\vec{Q},E)P(\vec{Q})\left\Vert F(\vec{Q},E)\right\Vert ^{2}\label{mathed:I_meas-I_back}\end{equation}
where $I_{bgd}$ is the background intensity, including fluorescence
or diffuse scattering, $r_{0}$ is the classic electron radius, $S$
is a scale factor, $A$ is the absorption correction, $D$ is the
detector efficiency, $L$ and $P$ are the Lorentz and polarization
factors for the Thomson scattering, and $\frac{1}{sin\theta }$ takes
into account the variation of the size of the beam footprint on the
sample surface when changing the incident angle. The $D$ factor takes
into account the whole detection set-up, comprising the detector efficiencies
and the absorption all the way from the monitor to the diffraction
detector. It can be obtained by making a base-line measurement of
the direct beam as a function of energy. It turns out that the energy
dependence of this factor is linear inside the energy range of interest
and so $D$ may be fitted to the DAFS data with a straight line ($D=m(\Delta E+1)$),
where \emph{m} is the only adjustable parameter, $\Delta E=E-E_{0}$,
and $E_{0}$ is energy at the edge. In that case, care should be taken
to measure the DAFS spectrum far enough from the absorption edge up
to the point where anomalous effect is negligible, otherwise the \emph{m}
parameter will be correlated to crystallographic phase $\Delta \varphi $.
It is always recommendable to measure a base-line, then the $D$ factor
is readily recovered. For a rotation scan, i.e. with the rotation
axis perpendicular to the plane of incidence (containing $\overrightarrow{k}$
and $\overrightarrow{k}^{\prime }$), $L=\frac{\lambda ^{3}}{\sin 2\theta }$,
the polarisation correction for the Thomson scattering is given by
the dot product $\hat{\varepsilon }.\hat{\varepsilon }^{\prime }$,
where $\hat{\varepsilon }$ and $\hat{\varepsilon }^{\prime }$ are
the polarization vectors of the incoming and outgoing photons, respectively.
If the electric field is perpendicular to the plane of incidence ($\sigma -\sigma $
scattering), the polarization factor is \emph{$P=p+(1-p)\cos ^{2}2\theta $},
where p is the $\sigma $-polarization rate. It is $P=p\cos ^{2}2\theta +(1-p)$
when the electric field is in the plane ($\pi -\pi $ scattering).
At a bending magnet, the polarization of the incoming beam is almost
entirely linear and \emph{p} is close to unity (95\% at the beamline
D2AM). For different geometries see the international table for crystallography
\cite{TableCritallo}.

Normalisation of the EDAFS oscillations can readily be obtained by
mutiplying the extracted signal $\frac{\frac{I_{bs}}{A(\vec{Q},E)}-I_{s}}{I_{s}}$
(where $I_{s}$ is the diffracted intensity without oscillations)
by \cite{Proietti99}:

\begin{equation}
S_{D}=\left\{ [\cos (\phi _{T}-\phi _{A})+\beta f_{0A}^{\prime }]^{2}+[\sin (\phi _{T}-\phi _{A})+\beta f_{0A}^{\prime \prime }]^{2}\right\} ^{\frac{1}{2}}/(2\beta \Delta f_{0A}^{\prime \prime })\label{mathed:SD}\end{equation}

\subsection{Absorption correction}

\label{sub:Absorption-correction}Absorption is a concern for the
DAFS spectroscopy because it introduces significant structure at the
edge that is strongly correlated to $\Delta $$\varphi $ and can
introduces distortions to the Extended-DAFS oscillations. Therefore
it is important to evaluate the effect of absorption before measuring
a DAFS spectrum. In this section we discuss on data reduction and
show how to correct for absorption and to minimize its effect by proper
measurement strategy. We recall the general expression which represents
the absorbed intensity for an incident beam impinging on a flat sample
of thickness \emph{t}, in reflection geometry, with the diffraction
vector perpendicular to the surface (symmetric reflection):

\begin{equation}
A(\vec{Q},E)=\int _{0}^{t}e^{-2\mu z/sin\theta _{B}}dz=\frac{1-e^{-2\mu t/sin\theta _{B}}}{2\mu /sin\theta _{B}}\label{mathed:A(Q,E)}\end{equation}

Where $\mu \left[m^{-1}\right]$ is the absorption coefficient, $\theta _{B}$
is the Bragg angle, and the factor 2 in the exponential takes into
account the absorption by both the incident and diffracted beams.
Equation \ref{mathed:A(Q,E)} applies only to symmetric reflection,
for other reflection geometries see the International Tables for Crystallography
\cite{TableCritallo}. If the film thickness is small enough to have
$2\mu t/\sin \theta _{B}$$\ll 1$, then $A\rightarrow t$, a constant.
For a bulk sample, $A=\frac{sin\theta _{B}}{2\mu }$, and the absorption
effect is strong. Correction to anomalous diffraction data affected
by strong absorption has been studied, for instance, by Vacinova et
al. \cite{Vacinova95a}, Bos \cite{Bos}, Bernhoeft \cite{Bernhoeft99}.
In the case of nearly perfect crystal and strong reflection, Meyer
et al. \cite{Meyer03} have given a quantitative approach to correct
for secondary extinction. 

In principle, we need the true $\mu $ to calculate equation \ref{mathed:A(Q,E)}.
The true $\mu $, however, is not generally available. From equation
\ref{mathed:f''A(sigma)}, the absorption coefficient can be calculated
in the forward scattering limit :\begin{equation}
\mu [m^{-1}]=2r_{0}\frac{hc}{E}N_{T}f_{T}^{\prime \prime }(E)\left[1+\frac{N_{A}f_{A}^{\prime \prime }(E)}{N_{T}f_{T}^{\prime \prime }(E)}\right]\label{mathed:mu_0}\end{equation}
where \emph{}$\mu _{i}=N_{i}\times \sigma _{i}$, \emph{$N_{T}$ and
$N_{A}$} are the number of non resonant and resonant atoms per $m^{3}$,
respectively ; $f_{T}^{\prime \prime }$ and $f_{A}^{\prime \prime }$
are in electron units.

It can be also expressed as:

\begin{equation}
\mu [cm^{-1}]=69876.576\times \frac{1}{E[keV]\times V[\textrm{Å}^{\textrm{3}}]}\times N_{T}f_{T}^{\prime \prime }(E)\left[1+\frac{N_{A}f_{A}^{\prime \prime }(E)}{N_{T}f_{T}^{\prime \prime }(E)}\right]\label{mathed:mu_1}\end{equation}
 where \emph{$N_{T}$} and \emph{$N_{A}$} are the number of non resonant
and resonant atoms in the unit cell, respectiveley, \emph{E} the energy
of the incident beam in keV and \emph{V} the unit cell volume in $\textrm{Å}^{\textrm{3}}$.
Over a typically measured energy range, $f_{T}^{\prime \prime }(E)$
has a weak linear dependence on \emph{E} and can be expressed in equations
\ref{mathed:mu_0} and \ref{mathed:mu_1} as $f_{T}^{\prime \prime }(E)=f_{T}^{\prime \prime }(E_{0})(m\Delta E+1)$,
where m is an adjustable parameter. If $N_{A}f_{A}^{\prime \prime }\ll N_{T}f_{T}^{\prime \prime }$
in equation \ref{mathed:mu_1}, then the absorption coefficient is
a constant.

The imaginary part $f_{A}^{\prime \prime }(E)$ in equation \ref{mathed:mu_0},
can be calculated either from the fluorescence spectrum, using for
instance the software DIFFKK \cite{Cross98b} (note that in case of
strong absorption, the fluorescence spectrum is distorted by self-absorption
and corrections must be applied \cite{Troger92}), or from theoretical
values. An experimental $A(\overrightarrow{Q},E)$ can be obtained
at a given $\overrightarrow{Q}$ by measuring the DAFS spectrum of
a Bragg peak which is not sensitive to the anomalous atoms (i.e. the
substrate or the buffer peaks or, if available, a reflection for which
anomalous atoms have a negligible contribution). Then, experimental
$N_{A}$, $N_{T}f_{T}^{\prime \prime }(E_{0})$ and $m$ can be recovered
using eq. \ref{mathed:A(Q,E)} to fit the experimental $A(\overrightarrow{Q},E)$,
where $N_{A}$, $N_{T}f_{T}^{\prime \prime }(E_{0})$ and $m$ are
adjustable variables. In this way, one can recover the true absorption
coefficient $\mu $ from the fluorescence signal spectrum and the
absorption jump at the edge in a DAFS spectrum from an experimental
$A(\overrightarrow{Q},E)$. Note that when using the substrate or
buffer peaks, a constant equal to $\frac{1}{2\mu _{sub.}/sin(\theta )}$
must be added to thin film absorption correction, where $\mu _{sub.}$
is the substrate or buffer absorption coefficient.

\begin{figure}[htbp]
\begin{center}\includegraphics[  width=1.0\textwidth,
  keepaspectratio]{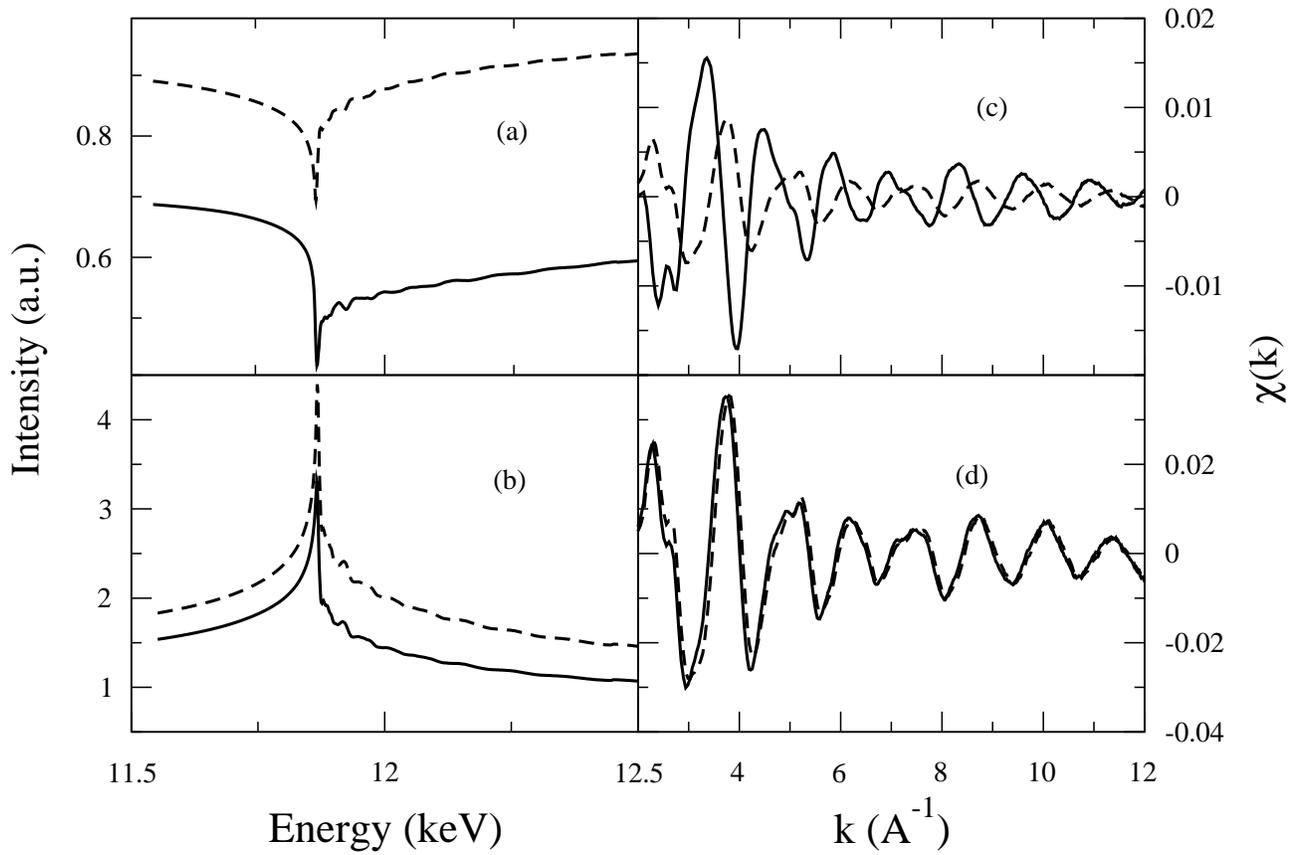}\end{center}

\caption{\label{fig:absorption_effect} left panel : (a) DAFS spectra of \emph{InAs}
(004) reflection, multiplied by $A(\vec{Q},E)/t$ calculated with
$t=0$ ($A(\vec{Q},E)=1)$ - \emph{dashed line} - and $t=2.8\mu m$
(one tenth of an absorption lenght) - \emph{solid line} - (b) same
as (a) for \emph{InAs} (006) reflection ; The total x-ray path length
inside the sample, $\frac{2t}{sin(\theta _{B})}$, is about $19\mu m$
for (004) and $11\mu m$ for (006). Right panel : (c) Extended DAFS
oscillations of \emph{InAs} (004) reflection are strongly affected
by absorption, whereas, (d) Extended DAFS oscillations of InAs (006)
reflection are not.}
\end{figure}

Although absorption effects are often small when measuring thin films,
this does not mean that there is no way to measure bulk samples or
that absorption is always negligible for thin samples. The relevant
quantity is the absorption jump amplitude, given to first order by
$2\mu t/sin\theta _{B}$, relative to the anomalous effect. Equation
\ref{mathed:F0_square} shows that the higher the parameter $\beta (\overrightarrow{Q})$
the higher the relative anomalous amplitude. High $\beta $ values
correspond to high $\left\Vert \alpha _{A}\right\Vert $ (ex : all
anomalous atoms are in phase) and small $\left\Vert F_{T}\right\Vert $.
Almost systematically, \emph{weak} reflections fulfill that criteria
and occur when resonant and non resonant atoms scatter out of phase.
As an example, take two reflections of the zincblende-like InAs compound:
the strong (004=4n) for which In and As scatter in phase, and the
weak (006=4n+2) for which In and As scatter out of phase. The values
for $\beta $, $\left\Vert \alpha _{A}\right\Vert $ and $\left\Vert F_{T}\right\Vert $
are reported in table \ref{table:InAs_absorption_effect} at the As
K-edge. Figure \ref{fig:absorption_effect} shows the (004) and (006)
DAFS spectra calculated by using experimental $f_{As}^{\prime }$
and $f_{As}^{\prime \prime }$ obtained from a XAFS spectrum of bulk
InAs. The square modulus of the structure factor has been multiplied
by $A(\overrightarrow{Q},E)/t$ calculated for a sample thickness
\emph{t} of one tenth of an absorption lenght at the edge ($t=\frac{0.1}{\Delta \mu }\approx 2.8\mu m$,
the variation $\Delta f_{As}^{\prime \prime }$ of $f_{As}^{\prime \prime }$
at the edge is equal to 3.4). The (004) DAFS spectrum clearly exhibits
an absorption like shape, whereas the (006) reflection is much less
affected. 

\begin{table}[htbp]

\caption{\label{table:InAs_absorption_effect}$\left\Vert \alpha _{A}\right\Vert $,
$\left\Vert F_{T}\right\Vert $, $\beta $ and $\Delta \varphi $
values for the (004) and (006) reflections of \emph{InAs} at the As
K-edge. Calculations are done at the As K-edge (11.867 keV) and do
not take into account the Debye-Waller factors (M=0). }

\begin{center}

\begin{tabular}{|c|c|c|c|c|}
\hline 
&
$\left\Vert \alpha _{A}\right\Vert /4$&
$\left\Vert F_{T}\right\Vert /4$&
$\beta $&
$\Delta \varphi $\\
\hline 
(004)&
1&
55.06&
0.018&
2.6°\\
\hline 
(006)&
1&
8.4&
0.12&
-162.4°\\
\hline
\end{tabular}

\end{center}
\end{table}

We want also to point out, without entering in details (for EDAFS
data analysis see reference \cite{Proietti99}) that a proper choice
of the Bragg reflection can also help to minimize distortions to the
Extended-DAFS oscillations induced by absorption. The structure factor
$F(\vec{Q},E)$ can be split into its smooth and oscillatory parts,
as in equation 5 in reference \cite{Proietti99}. The oscillatory
behaviour of $A(\overrightarrow{Q},E)$ propagates into the DAFS oscillations
leading to distortions whose strength depends on the relative amplitude
of the absorption oscillations compared to the amplitude of the DAFS
oscillations. To first order $A(\overrightarrow{Q},E)/t\approx 1-2\mu t/sin\theta _{B}$,
then the relative oscillations amplitude of $A(\overrightarrow{Q},E)$
scales as $2\mu _{0}t/sin\theta _{B}$, where $\mu =\mu _{0}(1+\chi ^{\prime \prime })$.
We observe in figure \ref{fig:absorption_effect}d that the distortion
of the Extended DAFS is negligible for the (006) Bragg reflection
whereas it is strong for the (004), shown in figure \ref{fig:absorption_effect}c.
It has to be noted that amplitude contrast of the DAFS oscillations
at first order is weighted \cite{Proietti99} by $\frac{\left\Vert \alpha _{A}\right\Vert }{\left\Vert F_{0}\right\Vert }$and
at the second order by$\frac{\left\Vert \alpha _{A}\right\Vert ^{2}}{\left\Vert F_{0}\right\Vert ^{2}}$,
i.e. not weighted by the ratio $\beta =\frac{\left\Vert \alpha _{A}\right\Vert }{\left\Vert F_{T}\right\Vert }$
as for the amplitude of the anomalous effect. Also, with superlattice
or very weak reflections, we could have a situation in which, despite
a modest anomalous effect ($\beta \ll 1)$, oscillations show large
relative amplitude due to chemical shifts or polarization effect (ATS)
that makes anomalous sites inequivalent \cite{Cross98b,Toda98}. Forbidden
reflections are a limiting case where $\left\Vert F_{0}\right\Vert =\left\Vert \alpha _{A}\right\Vert =0$
and only DAFS oscillations due to ATS contribute at second order to
the signal. This is by far the most favorable case where relative
absorption correction is the weakest, indeed, forbidden reflections
of bulk samples (for which absorption is strong) can be measured and
corrected ($A(\overrightarrow{Q},E)=1/\mu $) \cite{Garcia00,Renevier01}.

\begin{figure}[htbp]
\begin{center}\subfigure[]{\includegraphics[  width=0.49\textwidth,
  keepaspectratio]{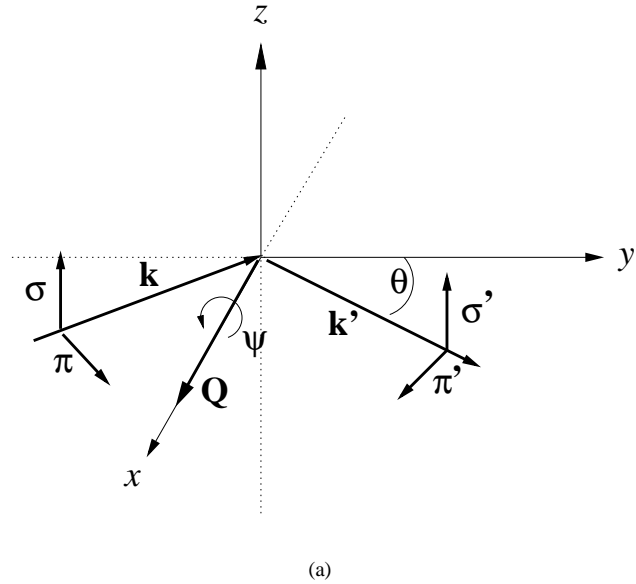}}\end{center}

\caption{\label{fig:scatt_geom_5.28} Schematic representation of the $\sigma -\pi $
and $\sigma -\sigma $ scattering geometries. (O,x,y) is the plane
of incidence in the schematic representation.}
\end{figure}

Finally, we want to mention how the absorption correction should be
taken into account in case of anisotropy of the scattering factor.
As stated in section \ref{sec:Elementary-background}, absorption
in non-cubic single crystals is not isotropic and depends on the polarization
vector direction of the incoming beam with respect to the principal
axis of the crystallographic point group. For instance, the absorption
cross section does not depend on polarization in a cubic crystal whereas
it does in a tetragonal one. Similarly, anomalous scattering is not
a scalar quantity, it depends on the polarization vector directions
of the incoming and outgoing beams with respect to the principal axis
of the point group symmetry of the resonant atoms (ATS). After summation
over all resonant atoms A, the ATS may produce scattering in the $\sigma -\sigma $
and/or $\sigma -\pi $ channels. Strictly speaking, this means that
the $\sigma -\sigma $ scattering only (incoming and outgoing polarization
vectors perpendicular to $\vec{Q}$), can be corrected with experimental
absorption data measured either in transmission or fluorescence mode,
provided that diffraction and absorption be measured with the polarization
of the incident beam in equivalent direction \cite{Cross98b} (see
figure \ref{fig:scatt_geom_5.28} for a schematic representation of
both scattering channels). Regarding $\sigma -\pi $ and $\pi -\pi $
scattering, there is no easy experimental solution, one can measure
two absorption spectra: one with the polarization vector parallel
to the polarization vector of the incoming beam and the other parallel
to that of the outgoing beam. The $\sigma -\pi $ scattering, that
comes from the off diagonal terms of $\left[D\right]$, has to be
checked only for forbidden or very weak reflections, when the diagonal
contribution almost cancel out. We want to note that care must be
taken with powder samples, since the absorption does not depend on
polarization but the DAFS does \cite{Bos}. 

\section{Concluding Remarks}

The aim of this paper is to give an up-to-date, general overview of
the technical aspects of DAFS spectroscopy. We believe that this technique
is underdeveloped compared with XAFS, despite providing significant
measurement capabilities unavailable to XAFS. Along with chemical
selectivity, DAFS offers spatial and site selectivity, allowing for
the application of well-established methods of XAFS data analysis
to new classes of problems. In this sense, we direct this paper to
the XAFS community. We hope that we have demonstrated that, despite
of the relative complexity of a DAFS measurement compared to XAFS,
significant progress has been made, making DAFS easier and more accessible.

We have thoroughly described the experimental set-up, paying attention
to the requirements of the design of the mirrors and monochromator
as well as to optical and sample alignment. These experimental practices
are essential to the collection of high quality data. We have also
discussed issues of detector design and mounting relevant to high
quality data collection. We use silicon photodiodes with a large dynamic
range able to cope with high x-ray fluxes and ensuring excellent signal-to-noise
on both the $I_{0}$ monitor and the diffraction detector comparable
with XAFS measurements. We also noted the importance of obtaining
a perfect monitor corrected $I/I_{0}$ normalization along the whole
energy scan.

Among the possible experimental schemes to perform energy-scan diffraction,
we have focused on the maximum intensity measurement procedure. It
has been dramatically improved at BM2-D2AM by developing a feedback
system that has proved to be very efficient in reducing measurement
time and increasing signal-to noise ratio. A further improvement to
the DAFS technique at BM2-D2AM has been the development of q-DAFS
combined with the feedback system. It allows collection of a DAFS
spectrum in a few minutes. We also treated the subject of absorption
correction, a major concern in data reduction. We have shown how to
use a fluorescence spectrum plus the jump at the edge of $A(\vec{Q},E)$
to recover the true $\mu $ spectrum and calculate the absorption
correction factor in a reliable way.

Finally, this paper provides information about the state of the art
of the DAFS technique at the beamline D2AM at the ESRF. We have tried
to make the application of the DAFS spectroscopy more attractive to
a wider section of the scientific community.

\section*{Acknowledgements}

We acknowledge the French CRG for granting beam time and support to
developping the DAFS spectroscopy at the beamline BM2-D2AM. One author
(BR) is pleased to acknowledge post-doctoral fellowship support from
the {}``Centre National pour la Recherche Scientifique''. We are
grateful to Y. Joly, E. Lorenzo, G. Subias and J. Garcia-Ruiz for
many helpful discussions and to the PhD students J. Vacinova, S. Bos,
V. Favre-Nicolin for their contributions to the developpement of the
DAFS spectroscopy. A special thanks to all users who have manifested
interest in DAFS. 

\bibliographystyle{apalike}
\bibliography{/home/renevier/litterature/BiblioTex/dafs,/home/renevier/litterature/BiblioTex/diffraction,/home/renevier/litterature/BiblioTex/exafs,/home/renevier/litterature/BiblioTex/exper,/home/renevier/litterature/BiblioTex/nano,/home/renevier/litterature/BiblioTex/navo,/home/renevier/litterature/BiblioTex/references}

\listoffigures

\end{document}